\documentclass[twocolumn,nofootinbib,amsmath,prd,aps,superscriptaddress,tightenlines,preprintnumbers]{revtex4}
\usepackage{graphicx}
\usepackage{epstopdf}
\usepackage{amsfonts}
\usepackage{bm}
\usepackage{epsfig}
\usepackage{xspace}
\usepackage[usenames]{color}
\def\bea{\begin{eqnarray}}
\def\eea{\end{eqnarray}}
\def\ba{\begin{eqnarray}}
\def\ea{\end{eqnarray}}
\def\be{\begin{equation}}
\def\ee{\end{equation}}
\def\beq{\begin{equation}}
\def\eeq{\end{equation}}

\newcommand{\lsim}{\mathrel{\rlap{\lower4pt\hbox{\hskip1pt$\sim$}}
    \raise1pt\hbox{$<$}}}         
\newcommand{\gsim}{\mathrel{\rlap{\lower4pt\hbox{\hskip1pt$\sim$}}
    \raise1pt\hbox{$>$}}}         

\newcommand{\leftrightarrowraised}{\mathrel{\rlap{\lower-0pt\hbox{\hskip1pt$\partial$}}
    \raise6 pt\hbox{$\leftrightarrow$}}}

\begin{document}

\newcount\hour \newcount\minute
\hour=\time \divide \hour by 60
\minute=\time
\count99=\hour \multiply \count99 by -60 \advance \minute by \count99
\newcommand{\mydate}{\ \today \ - \number\hour :00}

\title{Size of direct CP violation in singly Cabibbo-suppressed $D$ decays}

\def\Cincy{Department of Physics, University of Cincinnati, Cincinnati, Ohio 45221,USA}

\author{Joachim Brod}
\email[Electronic address:]{brodjm@ucmail.uc.edu} 
\affiliation{\Cincy}

\author{Alexander L. Kagan}
\email[Electronic address:]{kaganalexander@gmail.com} 
\affiliation{\Cincy}

\author{Jure Zupan}
\email[Electronic address:]{jure.zupan@cern.ch}
\email[On leave of absence from Josef Stefan Institute and
U. of Ljubljana, Ljubljana, Slovenia.]{}
\affiliation{\Cincy}

\begin{abstract}
The first experimental evidence for direct CP violation in charm-quark decays has recently been presented by the LHCb collaboration in the difference between the $D \to K^+ K^- $ and $D\to  \pi^+ \pi^- $ time-integrated CP asymmetries. We estimate the
size of the effects that can be expected within the Standard Model and find that at leading order in $1/m_c$
they are an order of magnitude smaller.
However, tree-level annihilation type amplitudes are 
known to be large experimentally. This implies that 
certain formally $1/m_c$-suppressed penguin amplitudes could plausibly account for 
the LHCb measurement. Simultaneously, the flavor-breaking parts of these amplitudes could explain the large difference between the $D \to K^+ K^- $ and $D\to  \pi^+ \pi^- $ decay rates. 
\end{abstract}

\maketitle
\newpage

The LHCb collaboration has recently announced a measurement of the
difference between the time-integrated CP asymmetries in two singly
Cabibbo-supressed (SCS) $D$-meson decay modes, ${\cal A}_{CP}(D\to
K^+K^-) - {\cal A}_{CP}(D\to \pi^+\pi^-) = (-0.82 \pm 0.21 \pm
0.11)\%$~\cite{Aaij:2011in}.  In the standard model (SM), CP violation
(CPV) in $D$ decays is commonly expected to be very small. It has
often been stated that ``a measurement of CP violation in $D$ decays
would be a signal of new physics". If this statement holds under
further scrutiny, the LHCb measurement may well be the first evidence
for new physics (NP) at the LHC.  The measurement can be identified,
to excellent approximation, with the difference between the two direct
CP asymmetries 
\beq \Delta {\cal A}_{CP} \equiv {\cal A}^{\rm
  dir}_{K^+ K^- } - {\cal A}^{\rm dir}_{\pi^+ \pi^- }
\eeq 
(the contribution of indirect CPV, due to a small difference in the
cut on the $D^0$ proper decay time in the two decays, is constrained
to be $\lsim 0.03$\%).  The new world average for the direct CP
asymmetry difference becomes $\Delta {\cal A}_{CP} = (-0.67 \pm
0.16)\%$ \cite{HFAG}, after the inclusion of the most recent CDF,
BABAR and Belle measurements
\cite{Collaboration:2012qw,Aubert:2007if,Staric:2008rx}. We are
therefore motivated to examine the size of direct CPV in the SM.

The role of the third generation in $D$ meson decays and $D^0-\bar
D^0$ mixing is suppressed by small Cabibbo-Kobayashi-Maskawa (CKM)
matrix elements.  Thus, all observables are CP conserving to first
approximation.  More precisely, in the SM CPV in $D^0-\bar D^0$ mixing
arises at $\mathcal O(V_{cb} V_{ub}/V_{cs} V_{us})\sim 10^{-3}$.
Direct CPV in SCS decays is further parametrically suppressed to
$\mathcal O([V_{cb} V_{ub}/V_{cs} V_{us}]\alpha_s/\pi)\sim 10^{-4}$,
since it comes from the interference of the tree and penguin
amplitudes.  Does this naive scaling estimate for $\Delta {\cal
  A}_{CP}$ mean that NP has been discovered? It is interesting that
despite severe constraints from other FCNC processes, $D-\bar D$
mixing in particular, it is possible to attribute $\Delta {\cal
  A}_{CP}$ entirely to NP \cite{Grossman:2006jg}.  As far as the SM
predictions are concerned, in \cite{Golden:1989qx} it was argued that
large CP asymmetries can be expected in SCS $D$ decays, while in
\cite{Buccella:1994nf} small CP asymmetries were obtained. Since these
early attempts, much more experimental information on $D$ decays has
become available. In this paper we address the following question: are
there any indications in the data that $\Delta {\cal A}_{CP}$ is due
to SM contributions that are enhanced well above the naive scaling
estimate?

We will show that there is a possible dynamical explanation for
$\Delta {\cal A}_{CP}$ in the SM.  A requirement 
for it to be viable is that 
the two matrix elements of the ``tree" operators in the effective weak
Hamiltonian, 
\beq\label{contracted}
 \langle K^+ K^-|(\bar d d)_{V-A} (\bar u c)_{V-A} |D^0\rangle,
\eeq
and
\beq\label{uncontracted}
\langle K^+ K^-|(\bar s s)_{V-A} (\bar u c)_{V-A} |D^0\rangle,
\eeq
are of similar magnitude.
(For simplicity we show only one of the two possible color structures
in \eqref{contracted} and \eqref{uncontracted}.) 
The analogous requirement applies to the matrix elements obtained via
the replacements $K\to \pi$ and $\ s\leftrightarrow d$. The matrix
element \eqref{contracted} only contains contributions in which the
$\bar d$ and $d$ fermion fields are contracted in the local
operator. The matrix element \eqref{uncontracted} receives contracted
as well as uncontracted contributions, i.e. the contributions where
the $\bar s$ and $s$ fermion fields are contracted with the final
state kaons.  Our requirement is therefore equivalent to saying that
the contracted contributions need to be enhanced over the uncontracted
ones.   

We will give two indications that this is the case.
First, assuming that the two matrix elements \eqref{contracted} and \eqref{uncontracted} 
differ by ${\mathcal O} (f_K/f_\pi-1 )\sim 0.2$, a quantity identified
with nominal SU(3) breaking, it is possible to {\it simultaneously}
explain both the large ratio
$Br(D\to K^+K^-)/Br(D\to \pi^+\pi^-)\simeq3.3$ and the enhancement of $\Delta {\cal A}_{CP}$. Second, using a one-gluon exchange approximation, we identify formally power-suppressed contracted contributions in \eqref{contracted} and \eqref{uncontracted}  that indeed show dynamical enhacement.
It is important to note that the enhancement of contracted over uncontracted contributions in \eqref{contracted} and \eqref{uncontracted} is in principle measurable on the lattice \cite{Hansen:2012tf}. Therefore, the SM explanation of large $\Delta {\cal A}_{CP}$ can be definitively checked in the future.

We begin with a discussion of enhanced contractions in the one-gluon exchange approximation.
The essential ingredients are:  (i) $1/N_c$ counting; 
(ii)  $D$ branching ratio data which shows that certain formally $1/m_c$ power-suppressed amplitudes are of same order as their leading $(1/m_c)^0$ counterparts; 
(iii) translation of this breakdown of the $1/m_c$ expansion to the penguin contraction amplitudes, in the approximation of a hard gluon exchange;
(iv) a perturbative estimate of the related ``effective Wilson coefficients".

The starting point in estimating
${\cal A}^{\rm dir}_{K^+ K^- }$ and ${\cal A}^{\rm dir}_{\pi^+ \pi^-
}$ is the effective weak $\Delta C=1$
Hamiltonian $H_{\rm eff} $, evaluated at scales $\mu \sim m_c $.  It is obtained by integrating out the $W$ boson
and running at NLO down to $\mu \approx m_b$, where the
$b$-quark is integrated out using NLO matching. Finally, running down to $\mu \sim m_c $
yields~\cite{Buchalla:1995vs}
\begin{equation}
\begin{split}
H_{\rm eff}^{\Delta C=1} &= \frac{G_F}{\sqrt{2}} \Big[\sum_{p=d,s} 
V_{cp}^* V_{up}  \left(C_1 Q_1^p + C_2 Q_2^p \right) \\
&-V_{cb}^* V_{ub} \sum_{i=3}^{6} 
C_i Q_i + C_{8g} Q_{8 g} \Big]
+ {\rm H.c.}
\end{split}
\end{equation}
The ``tree" operators are $ Q_1^p=(\bar p c)_{V-A}(\bar u p)_{V-A},
Q_2^p=(\bar p_\alpha c_\beta)_{V-A} (\bar u_\beta p_\alpha)_{V-A} $
with summation over color indices $\alpha,\beta$ understood. The
penguin operators are $Q_{3,5}=(\bar u c)_{V-A} \sum_q (\bar qq)_{V\mp
  A}$, $Q_{4,6}=(\bar u_\alpha c_\beta)_{V-A} \sum_q (\bar q_\beta
q_\alpha)_{V\mp A}$ and $Q_{8g} = -\frac{g_s}{8\pi^2}\, m_c \bar u
\,\sigma_{\mu\nu}(1+\gamma_5) G^{\mu\nu} c$, where the summation is
over the light quark species, $q=u,d,s$. The Wilson coefficients $C_i
(\mu) $ are evaluated at $\mu \sim m_c $. We find $C_1=1.21$,
$C_2=-0.41$, $C_3=0.02$, $C_4=-0.04$, $C_5=0.01$, $C_6=-0.05$,
$C_{8g}=-0.06$. 

The CP-conjugate decay amplitudes for SCS $D \to f $ and $\bar D \to \bar f$ decays in the SM can be written as 
\beq
\begin{split}
A_f   (D \to f )  &=   A^T_{f} [1+r_f e^{i(\delta_f-\gamma)}],\\
\overline{A}_{\overline f}(\bar D \to \bar{f}  ) &=  A^T_{f}[1+r_f e^{i(\delta_f+\gamma)}],
\end{split}
\eeq
where $A^T_{f} $ is the dominant tree-level amplitude, taken to be real and positive by
convention, and $r_f =A_f^P/A_f^T$ is the relative magnitude of the subleading ``penguin" amplitude, $A^P_f$, which 
carries the CKM weak phase $\gamma=(67.3^{+4.2}_{-3.5})^\circ$~\cite{Charles:2011va}. The penguin amplitude is CKM suppressed by  ${\mathcal O}(V_{cb} V_{ub}/V_{cs} V_{us})$ compared to the tree-level amplitude and carries the relative strong phase
$\delta_f $. The direct CP asymmetry is then 
\beq\label{Adir.eq}
{\cal A}_f^{\rm dir}   \equiv { |A_f |^2  - | \bar A_{\bar f } |^2 \over  |A_f |^2  + | \bar A_{\bar f } |^2 }  = 2 r_f \sin \gamma \sin \delta_f,
\eeq
where we only kept the term linear in $r_f$. 

From the measured $D\to f$ branching ratios we can obtain information on the sizes of the tree-level amplitudes $A_f^T$. 
It is important to note that the data on SCS or Cabibbo favored (CF) decays implies that
formally $1/m_c$-suppressed tree-level amplitudes are of the same order
as the leading power contributions. For instance, 
the tree level CF decay $D^0 \to K^- \pi^+$, written in $SU(3)_F$ diagrammatic notation, is
\beq
\begin{split}
A^T_{K^+ \pi^- }&  =  V_{cs}^* V_{ud} ( T_{K\pi} + E_{K\pi}), 
\label{eq:Tamps}
\end{split}
\eeq 
and similarly for the tree-level amplitudes in $D^0 \to K^+ K^- $, $\pi^+ \pi^- $ with appropriately modified CKM elements.
Here $T_{f}$ are the ``color-allowed $W$-emission"
amplitudes, and $E_{f}$ are the ``$W$-exchange" annihilation amplitudes~\cite{Chau:1982da,Gronau:1994rj}.  
Equating the magnitudes of the total tree amplitudes to the measured
ones, one has $ A^{T}_{K^-  \pi^+} \simeq 2.5 ~{\rm keV}\sim A^T_{\pi^+\pi^-}/\lambda\sim A^T_{K^+K^-}/\lambda$, with $\lambda=|V_{us}|=0.22$.

At leading power and in naive factorization $T_f$ take the familiar form, e.g., $T_{\pi\pi } \propto F_{D \to \pi} f_{\pi}$, where $F_{D\to \pi}$ is  the form factor and $f_\pi$ the decay constant.  
The annihilation amplitudes
$E_{f}$  
are formally $1/m_c$ suppressed power corrections. However, SU(3) fits to the data yield ($f=KK,\, \pi\pi, K\pi$) \cite{Bhattacharya:2009ps, Cheng:2010ry} 
\beq 
E_{f} \sim   T_{f}\,~~{\rm and}~~E_{KK}\sim E_{\pi\pi} \sim E_{K\pi}\,.\label{appTE}
\eeq 
The first relation signals the breakdown of the $1/m_c$ expansion. 

The QCD penguin amplitudes are
\beq
A^{P}_{K^+ K^-} = -V_{cb}^* V_{ub}  P_{KK} ,~~~~A^P_{\pi^+ \pi^-} =-V_{cb}^* V_{ub}  P_{\pi\pi},
\eeq
yielding weak phases $-\gamma$ and $\pi -\gamma $ relative to the $D\to \pi\pi$ and $D\to KK$ tree amplitudes,  respectively, with 
$\sin\gamma \simeq 0.9$. This implies that the CP asymmetries in $D\to \pi\pi$ and $D\to KK$ have opposite signs.

First let us check whether the naive CKM scaling estimate $r_f\sim {\mathcal O}(10^{-4})$ is reproduced in the formal $m_c\to \infty$ limit.  Rough estimates for these leading-power penguin amplitudes are obtained using QCD factorization at NLO in $\alpha_s$ \cite{Beneke:1999br,Beneke:2001ev,Grossman:2006jg}, i.e.  in
naive factorization plus the $O(\alpha_s) $ corrections coming from down- and strange-quark loop penguin contractions, vertex corrections, and hard spectator interactions.  For the penguin to measured tree amplitude ratios 
\beq
   r^{\rm LP}_f \equiv \left|{A^P_f ({\rm leading~power}) \over A^T_f ({\rm exp}) }\right|, \label{LOr} 
   \eeq
we obtain $ r^{\rm LP}_{K^+ K^-} \approx (0.01 - 0.02)\, \%,~~~ r^{\rm
  LP}_{\pi^+ \pi^-} \approx (0.015 - 0.028)\%$, for a renormalization scale in the range $\mu \in [1\,{\rm GeV}, m_D]$, and down and strange quark masses in the penguin contraction loops varied in the ranges $m_d , m_s  \sim 0.1 - 0.4$ GeV.
This is consistent with, yet slightly bigger than, the naive expectation that $r_f \sim \mathcal O([V_{cb} V_{ub}/V_{cs} V_{us}]\alpha_s/\pi)\sim 10^{-4}$, based only on CKM factors and $\alpha_s$ scaling. 
Assuming ${\mathcal O}(1)$ strong phases $\delta_{f}$, the direct CP asymmetries satisfy ${\cal A}_f^{\rm dir} \sim 2 r_f$.
The $U$-spin relation ${\cal A}_{\pi\pi}^{\rm dir} = -{\cal A}_{KK}^{\rm dir}$ then leads to $\Delta {\cal A}_{CP}\sim 4 r_f$,
or $\Delta {\cal A}_{CP} = O(0.05\%-0.1\%)$,
an order of magnitude smaller than the measurement. 
However, already at leading power our estimates for the penguin amplitudes are dominated by the penguin contractions.

\begin{figure}[t]
\centering
\hbox{\hspace{-.25cm}$\begin{array}{ccc}
\includegraphics[width=4.0cm]{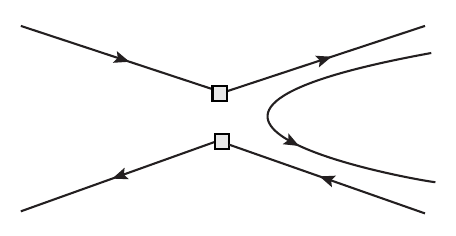} &
\includegraphics[width=4.0cm]{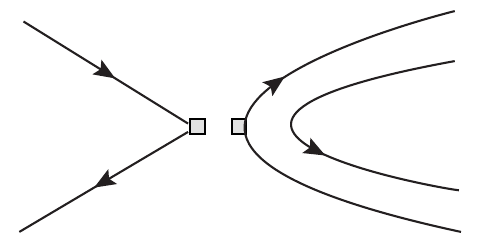}
\end{array}$}
\caption{Annihilation topologies. \label{fig:T} }
\end{figure}

Next, we turn to the QCD penguin power corrections. 
For concreteness we focus on the annihilation topology amplitudes, which in the tree amplitudes are seen to be of the same size as the leading power contributions, $T_f\sim E_f$. We consider two examples of $1/m_c$ penguin amplitudes for the final states $f=K^+K^-$ and $\pi^+\pi^-$,
\beq
\begin{split}\label{eq:amp1}
P_{f,1} =& {G_F \over \sqrt 2}\, C_6^{}  \langle f | \!-2\, (\bar u u )_{S+P} \otimes^A (\bar u  c )_{S-P} |D  \rangle,
\end{split}
\eeq
and
\beq
\begin{split}\label{eq:amp2}
P_{f,2} =& {G_F \over \sqrt 2}  \,(C_4^{} + C_6 ^{} ) \langle f | \,(\bar q_\alpha  q_\beta )_{V\pm A} \otimes^A (\bar u_\beta  c_\alpha )_{V-A} |D  \rangle,
\end{split}
\eeq
corresponding to insertions of the QCD penguin
operators in the left and right annihilation topology diagrams of Fig.~\ref{fig:T}, respectively. 
Summation over $q=u,d,s$ is understood.  
The annihilation product $J_1 \otimes^A J_2 $ means that $J_2$ destroys the $D$ meson, and $J_1$ creates the final state.  
We will estimate the non-perturbative matrix elements in~\eqref{eq:amp1} and \eqref{eq:amp2} from the sizes of 
$E_{f}$ in \eqref{eq:Tamps}, see below.

\begin{figure}[t]
\centering
\hbox{\hspace{-.25cm}$\begin{array}{ccc}
\includegraphics[width=4.0cm]{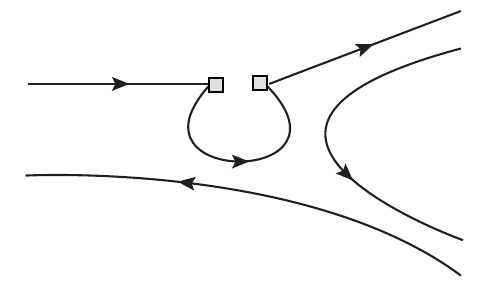} &
\includegraphics[width=4.0cm]{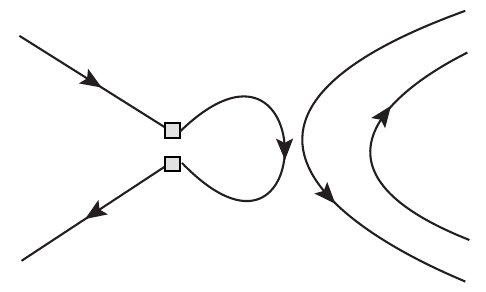}
\end{array}$}
\caption{Penguin contraction topologies. \label{fig:P}}
\end{figure}

The penguin amplitudes $P_{f,1}$ and $P_{f,2}$ also receive contributions from penguin contractions of the tree operator $Q_1$, shown in Fig.~\ref{fig:P} (left) and Fig.~\ref{fig:P} (right), respectively.  These contributions cancel
the $\log\mu $ scale dependence of the Wilson coefficients in \eqref{eq:amp1} and \eqref{eq:amp2}.  
In the partonic picture the cancelation in $P_{f,1}$ ($P_{f,2}$) is associated with the exchange of a single gluon between the 
$s$ and $d$-quark loops
and the spectator quark (outgoing $q\bar q$ pair), but an arbitrary number of partons otherwise exchanged between the external legs.  This leads to effective coefficients for $C_{4,6}$ 
that depend on the gluon's virtuality $q^2$, 
\beq
\begin{split}\label{eq:c6eff}
C_{6 \,(4)}^{\rm eff} &\big( \mu, \tfrac{q^2}{m_c^2} \big) =  C_{6 \,(4)} (\mu ) + C_1 (\mu )\, {\alpha_s (\mu )\over 2 \pi}\\
 &\times\Big[{1\over 6} +
  {1\over 3}
 \log\left({m_c \over \mu} \right) -{1\over 8}
 G\Big(\tfrac{m^2_s}{m_c^2 }, \tfrac{m^2_d}{m_c^2 } ,
 \tfrac{q^2}{m_c^2}\Big)\Big] ,
 \end{split}
\eeq
where $G(s_1,s_2,x) = (G(s_1,x)+G(s_2,x))/2$ is an infrared finite quantity.  At one loop, $G(s,x)$ is 
defined, e.g., in~\cite{Beneke:2001ev}.  

In order to obtain rough values for the penguin contraction amplitudes we make two approximations:
(i) we use the partonic quantity $G$ as an estimator for the
underlying hadronic interactions, for instance, 
final state interactions,
(ii) we take $G$ at a fixed value of $q^2 = \hat{q}^2$.
With these approximations we can replace 
$C_{4,6}$ with $C_{4,6}^{\rm eff} $ in \eqref{eq:amp1} and \eqref{eq:amp2}.
Order of magnitude estimates of the annihilation-type matrix elements in
\eqref{eq:amp1}, \eqref{eq:amp2} are obtained by
appealing to the ``$W$-exchange" annihilation amplitudes $E_f$ in~\eqref{eq:Tamps}.
For instance, for $D^0\to K^+\pi^-$, we have
\beq
\begin{split}
E_{K\pi}  = &  {G_F \over \sqrt 2}  C_1  \langle K^+ \pi^- |
(\bar s_{\alpha} d_{\beta} )_{V- A} \otimes^A (\bar u_\beta
c_\alpha)_{V-A}| D  \rangle  \,, 
\label{Ef}
\end{split}
\eeq
neglecting a smaller contribution proportional to $C_2\sim -C_1/3$.
Note that the $E_f$ enter at ${\mathcal O}(1/N_c)$.

We are now able to study the ratios $r_f$ in Eq.  \eqref{LOr} beyond leading power,
\beq
   r^{\rm PC}_f \equiv \left|{A^P_f ({\rm power~correction}) \over A^T_f ({\rm exp}) }\right|\,,~~f= K^+ K^- \,,\pi^+ \pi^- ,\label{PCr} 
   \eeq 
 by assuming $N_c $ counting, and ignoring differences due to the
 different chirality structures, which should be 
small in $D \to PP$ decays. 
We thus equate the operator matrix element \eqref{Ef} with the matrix
elements of the operators in \eqref{eq:amp1} and \eqref{eq:amp2},
taking into account that $P_{f,1}$ is ${\mathcal O}(1)$ and $P_{f,2}$
is ${\mathcal O}(1/N_c)$.
We use the experimentally supported relations $E_{f} \sim T_{f}\sim A_f^T({\rm exp})$, cf. Eq. \eqref{appTE}.  Finally, to arrive at 
numerical estimates 
we set $E_f= A_f^T({\rm exp})$ and use the experimental observation $A_{f}^T \sim A^T_{K\pi} \theta_c $ to obtain 
\begin{align}\label{eq:rf1}
r_{f,1}&\equiv \left|\frac{A^P_{f,1}}{A^T_{f}}\right|\sim 2
N_c |V_{cb} V_{ub} C_6^{\rm eff}|/(C_1 \sin\theta_c),\\
r_{f,2}&\equiv \left|\frac{A^P_{f,2}}{A^T_{f}}\right|\sim  2 |V_{cb} V_{ub} (C_4^{\rm
  eff}+C_6^{\rm eff}) |/(C_1 \sin\theta_c).\label{eq:rf2}
\end{align} 

The above estimates for $r_f$ depend on $\hat q^2$ through the $G$ function. Taking $\hat q=m_c$, varying $\mu\in[1{\rm \, GeV}, m_D ]$ and setting $m_s = 0.3\,\text{GeV}$, $m_d =
0.1\,\text{GeV}$ we obtain $r_{f,1}\in [0.04, 0.08]\%$ and $r_{f,2}\in[0.03,0.04]\%$.
We see that the individual power corrections which we have considered
could be enhanced by a factor of a few relative to leading power. For example, for a renormalization scale $\mu=1$ GeV and taking
$\Delta {\cal A}_{CP}\sim 4 r_f$, we find $\Delta {\cal A}_{CP}\sim 0.3\% (0.2\%)$  for $P_{f,1} (P_{f,2})$. 
Assuming smaller constituent quark masses could give even larger effects. 

The enhancement of the penguin amplitude can be traced to (i) the breakdown of the $1/m_c$ expansion, via 
the comparison of the matrix elements in $P_{f,1}$, $P_{f,2}$ and $E_{K\pi}$, and (ii) the 
penguin contraction contributions, which dominate in  \eqref{eq:c6eff} via the $G$ function.

There are other contributions to the penguin amplitudes
at subleading order in $1/m_c $
which may be at least as large as those we have discussed.  An 
example is penguin contractions which correspond to emission of two or more gluons from the $s$ and $d$-quark loops 
in the partonic picture (these were discussed for $B\to K\pi$ in \cite{Duraisamy:2009kb}). 

We now address the other indication for enhanced penguin contractions.  Assume that this enhancement is of ${\mathcal O}(1/\epsilon)$, where $\epsilon\sim {\mathcal O}(20\%)$ is the typical size of SU(3) breaking.  Then
\begin{align}
\begin{split}
A(\bar D^0\to \pi^+\pi^-)=&V_{cd} V_{ud}^*\big(T+E+P_{\rm break}\big) \label{pipidiagrammatic}\\
&-V_{cb}^*V_{ub} P,
\end{split}
\\
\begin{split}
A(\bar D^0\to K^+K^-)=&V_{cs} V_{us}^*\big(T+E -P_{\rm break}\big) \label{KKdiagrammatic}\\
&-V_{cb}^*V_{ub} P,
\end{split}
\end{align}
where we only write down the leading contributions in $\epsilon$ counting for each product of CKM elements. The ${\mathcal O}(\epsilon^0)$ amplitudes $T$ and $E$ equal  $T_{K\pi}$ and $E_{K\pi}$ in \eqref{eq:Tamps} in the U-spin limit. The U-spin symmetric penguin amplitude
\beq
P\equiv\, \tfrac12 \langle f | \sum_{i=1,2} C_i (Q_i^d+Q_i^s) |\bar D^0\rangle\,,
\eeq
where $f= K^+K^-$ or $\pi^+\pi^-$, contains contracted contributions and is thus ${\mathcal O}(1/\epsilon)$. 
The U-spin breaking difference of the penguin contractions,
\beq  
\begin{split}
 P_{\rm break}\equiv \tfrac{1}{2}\sum_f \langle f| \sum_{i=1,2} C_i (Q_i^{ d}-Q_i^{s})|\bar D^0\rangle\,, \label{eq:t3}
\end{split}
\eeq
is part of the ${\mathcal O}(\epsilon\cdot 1/\epsilon)\sim {\mathcal O}(\epsilon^0)$ amplitude.
In particular, it gives ${\mathcal O}(1)$ corrections to the $A(\bar D^0\to \pi^+\pi^-)$ and $A(\bar D^0\to K^+K^-)$ amplitudes, explaining the large difference in the branching ratios (see also~\cite{Bhattacharya:2012ah}).

At the same time $P\sim P_{\rm break}/\epsilon$ is of the correct size to give 
\beq \label{randPovT}
\begin{split} 
r_{\pi^+\pi^-, K^+K^-}&\simeq
\left|\frac{V_{cb}V_{ub}}{V_{cs}V_{us}}\right|\cdot
\left|\frac{P}{T+E \pm P_{\rm break}}\right|\\
&\sim
\frac{|V_{cb}V_{ub}|}{|V_{cs}V_{us}|} \frac{1}{2 \epsilon}\sim 0.15 \%,
\end{split}
\eeq
for $\epsilon \sim 0.2$. For ${\mathcal O}(1)$ strong phases this gives an estimate $|\Delta {\mathcal A}_{CP}|\sim 4 r_f \sim 0.6 \%$ that is strikingly close to the world average \cite{HFAG} (for further details see \cite{Brod:2012ud}). This shows that for nominal SU(3) breaking of ${\mathcal O}(0.2)$, the enhancement of the contracted amplitudes can simultaneously explain the large difference of $Br(D\to K^+K^-)$ and $Br(D\to \pi^+\pi^-)$, and the SM enhancement of $\Delta {\cal A}_{CP}$ over the naive scaling estimate.

Given the measured value of $\Delta
{\cal A}_{CP}$, can we predict the orders of magnitude of direct CPV
in other SCS $D$ decays? For 
processes that only differ from $D^0\to K^+ K^-$ and $D^0\to \pi^+\pi^-$
by a change in the flavor of the spectator, we would generically expect
direct CP asymmetries of same order, 
both in the SM and in NP extensions that modify the
QCD penguin operators. Examples are $D^+\to K^+ \overline{K^0}$
and $D_s^+\to \pi^+ K^0$, which contain SM contributions from  
the penguin contractions of the type shown in Fig.~\ref{fig:P} (left) or $P_{f,1}$,
but not Fig.~\ref{fig:P} (right) or $P_{f,2}$.  
Interestingly, for $D\to K_S K_S$ the rate is due to the difference of $s$ and $d$ quark penguin contractions 
of the latter type, as well as the difference of exchange graphs $E$ with outgoing $s$ and $d$
quark pairs, while the SM penguin amplitude is due to the two sums.  Thus, a large direct 
CP asymmetry is possible in the SM for ${\cal O}(1)$ strong phases, 
${\cal A}^{\rm dir}_{K_S K_S }  \sim  {|V_{cb}V_{ub} / V_{cs}V_{us}|} {2}/{\epsilon} \sim 0.6\%$.
A study of QCD penguin amplitudes in SCS $D_{(s)}$ decays to vector and pseudoscalar pairs 
is also of interest.

It is of course not excluded that $\Delta {\cal A}_{CP}$ is (at least partly) due to NP. 
In fact, it is possible to prove the presence of NP in $\Delta I=3/2$ decays. 
For example, the $D^+\to \pi^+\pi^0$ amplitude does not receive a QCD penguin contribution.  It must vanish because $\pi^+\pi^0$ is an $I=2$ final state which cannot be reached from the initial state via the $\Delta I=1/2$ QCD penguin operators. It does, however, receive an electroweak penguin contribution which is $\alpha/\alpha_S$ suppressed in the SM. Thus, 
enhanced direct CPV in $D^+\to\pi^+\pi^0$, e.g., at the level of the observed $\Delta {\cal A}^{}_{CP}$, would be a signal for isospin violating NP \cite{Grossman:2012eb}.

In conclusion, we have shown that it is plausible that the SM accounts for the measured value of $\Delta {\cal A}_{CP}$.  
At leading order in $1/m_c$ we obtain $\Delta {\cal A}_{CP} = O(0.05\%-0.1\%)$,
an order of magnitude smaller than observed. However, we find that significant enhancements are possible for certain penguin contraction power corrections whose matrix elements can be estimated using information obtained from global $SU(3)$ fits to the measured $D$ decay branching ratios, utilizing $N_c$ counting 
for guidance. 
We reach the conclusion that the ratio of QCD penguin to tree amplitudes can be $r_f\sim 0.1 \% $. Using the $U$-spin relation ${\cal A}^{\rm dir}_{\pi^+ \pi^-} \simeq
- {\cal A}^{\rm dir}_{K^+ K^-}$  one has $\Delta {\cal
  A}_{CP}\simeq 2 {\cal A}^{\rm dir}  \simeq 4 r_f 
\sin\delta \sin\gamma $.
For  a large strong phase $\delta\sim O(1)$  this yields $\Delta {\cal A}_{CP}\sim 0.4\%$. 
This result is subject to large uncertainties due to (i) the extraction of the tree-level annihilation amplitudes $E_f$ from data, (ii) neglected contributions in $E_f$, (iii) the use of $N_c$ counting for penguin operator matrix elements, (iv) the modeling of $Q_1$ penguin contraction matrix elements, 
and  (v) additional penguin contractions, not associated with $\log\mu$ cancellations, that have not been included in our estimates.
A cumulative uncertainty of a factor of a few is reasonable.

Further support for the above SM explanation, where $\Delta {\cal A}_{CP}$ is due to the enhancement of contracted $Q_1$ operator contributions, is provided by the large difference $Br(D\to K^+K^-)\simeq 2.8 Br(D\to \pi^+\pi^-)$. It can be attributed to the SU(3) breaking part of the penguin contractions, if these are enhanced. This fixes the size of the penguin contractions and leads to the prediction $r_f\sim 0.2 \%$, in agreement with our semi-perturbative estimates, and most importantly, in agreement with the measured value of $\Delta {\cal A}_{CP}$ for strong phases of order ${\mathcal O}(1)$.

{\bf Acknowledgements:} We thank  Marco Gersabeck, Tim Gershon, Vladimir Gligorov, Zoltan Ligeti, Gilad Perez, Alexander Petrov, and Guy Wilkinson for discussions, and the CERN theory group for hospitality. A.~K.  and J.~B. are supported by DOE grant FG02-84-ER40153.
\\{}

{\bf Erratum:} Equation \eqref{eq:c6eff} contains a typographical error, where $ -{1\over 8}
 G\big(\tfrac{m^2_s}{m_c^2 }, \tfrac{m^2_d}{m_c^2 } ,
 \tfrac{q^2}{m_c^2}\big)$ should be replaced by $ -{1\over 4}
 G\big(\tfrac{m^2_s}{m_c^2 }, \tfrac{m^2_d}{m_c^2 } ,
 \tfrac{q^2}{m_c^2}\big)$. Additionally, the  numerical  ranges for $r_{f,1}$ and $r_{f,2}$ below \eqref{eq:rf2}  should be changed to $r_{f,1}\in [0.02,0.03]\%$  and $r_{f,2}\in [0.01,0.02]\%$. (We thank Uli Nierste and Stefan Schacht for correcting our evaluation of the $G$ function \cite{Nierste:2015zra}.)   Thus for a large strong phase $\delta\sim {\mathcal O}(1)$ this yields $\Delta {\cal A}_{CP} \lesssim0.2\%$ with a cumulative error of a factor of a few being reasonable.


\end{document}